\documentclass[twocolumn, switch]{article} % Method A for two-column formatting

\usepackage{preprint}

\usepackage{amsmath, amsthm, amssymb, amsfonts}

\usepackage{algorithmic}
\usepackage{algorithm}
\usepackage{array}
\usepackage[caption=false,font=normalsize,labelfont=sf,textfont=sf]{subfig}
\usepackage{textcomp}
\usepackage{stfloats}
\usepackage{url}
\usepackage{verbatim}
\usepackage{graphicx}

\usepackage[numbers,square]{natbib}
\usepackage[utf8]{inputenc}	% allow utf-8 input
\usepackage[T1]{fontenc}	% use 8-bit T1 fonts
\usepackage{xcolor}		% colors for hyperlinks
\usepackage{booktabs} 		% professional-quality tables
\usepackage{microtype}		% microtypography
\usepackage{lineno}		% Line numbers
\usepackage{float}			% Allows for figures within multicol

\usepackage{lipsum}		%  Filler text

\usepackage[hidelinks]{hyperref} %For e-mail href
\usepackage{float}
\usepackage{subfig}
\usepackage[nice]{nicefrac}
\usepackage{newfloat}
\DeclareFloatingEnvironment[name={Supplementary Figure}]{suppfigure}
\usepackage{sidecap}
\sidecaptionvpos{figure}{c}

\usepackage{titlesec}
\titlespacing\section{0pt}{12pt plus 3pt minus 3pt}{1pt plus 1pt minus 1pt}
\titlespacing\subsection{0pt}{10pt plus 3pt minus 3pt}{1pt plus 1pt minus 1pt}
\titlespacing\subsubsection{0pt}{8pt plus 3pt minus 3pt}{1pt plus 1pt minus 1pt}

\usepackage{tikz,xcolor,hyperref}

\definecolor{lime}{HTML}{A6CE39}
\DeclareRobustCommand{\orcidicon}{
	\begin{tikzpicture}
	\draw[lime, fill=lime] (0,0) 
	circle [radius=0.16] 
	node[white] {{\fontfamily{qag}\selectfont \tiny ID}};
	\draw[white, fill=white] (-0.0625,0.095) 
	circle [radius=0.007];
	\end{tikzpicture}
	\hspace{-2mm}
}
\foreach \x in {A, ..., Z}{\expandafter\xdef\csname orcid\x\endcsname{\noexpand\href{https://orcid.org/\csname orcidauthor\x\endcsname}
			{\noexpand\orcidicon}}
}
\title{Magnitude-Corrected and Time-Aligned Interpolation of Head-Related Transfer Functions}

\usepackage{authblk}

\author[1]{Johannes M. Arend\orcidA{}}
\author[2]{Christoph Pörschmann\orcidB{}}
\author[1]{Stefan Weinzierl\orcidC{}}
\author[1]{Fabian Brinkmann\orcidD{}}

\affil[1]{Audio Communication Group, Technical University of Berlin, Berlin 10587, Germany}
\affil[2]{Institute of Computer and Communication Technology, TH K\"oln - University of Applied Sciences, Cologne 50679, Germany}

\begin{document}

\twocolumn[ % Method A for two-column formatting
  \begin{@twocolumnfalse} % Method A for two-column formatting
  
\maketitle

\begin{abstract}
Head-related transfer functions (HRTFs) are essential for virtual acoustic realities, as they contain all cues for localizing sound sources in three-dimensional space. Acoustic measurements are one way to obtain high-quality HRTFs. To reduce measurement time, cost, and complexity of measurement systems, a promising approach is to capture only a few HRTFs on a sparse sampling grid and then upsample them to a dense HRTF set by interpolation. However, HRTF interpolation is challenging because small changes in source position can result in significant changes in the HRTF phase and magnitude response. Previous studies greatly improved the interpolation by time-aligning the HRTFs in preprocessing, but magnitude interpolation errors, especially in contralateral regions, remain a problem. Building upon the time-alignment approaches, we propose an additional post-interpolation magnitude correction derived from a frequency-smoothed HRTF representation. Employing all 96 individual simulated HRTF sets of the HUTUBS database, we show that the magnitude correction significantly reduces interpolation errors compared to state-of-the-art interpolation methods applying only time alignment. Our analysis shows that when upsampling very sparse HRTF sets, the subject-averaged magnitude error in the critical higher frequency range is up to 1.5\,dB lower when averaged over all directions and even up to 4\,dB lower in the contralateral region. As a result, the interaural level differences in the upsampled HRTFs are considerably improved. The proposed algorithm thus has the potential to further reduce the minimum number of HRTFs required for perceptually transparent interpolation.
\end{abstract}
%\keywords{First keyword \and Second keyword \and More} % (optional)
\vspace{0.35cm}

  \end{@twocolumnfalse} % Method A for two-column formatting
] % Method A for two-column formatting

%\begin{multicols}{2} % Method B for two-column formatting (doesn't play well with line numbers), comment out if using method A

%%%%%%%%%%%%%%%  Main text   %%%%%%%%%%%%%%%
% \linenumbers

\section{Introduction}

Head-related transfer functions (HRTFs) describe the direction-dependent acoustic filtering of incident sound due to the listener's morphology. They contain monaural and binaural cues that the auditory system uses for localization in the median plane (up/down) and horizontal plane (left/right), respectively. The monaural cues arise from spectral changes caused by the head, torso, and especially the pinnae, where sound is reflected and diffracted, resulting in direction-dependent patterns. The binaural cues arise from comparing both ear signals and are a combination of interaural time differences (ITDs), which mainly result from the distance between the ears, and interaural level differences (ILDs), which mainly result from the acoustic shadowing of the head~\cite{Blauert1996}.

HRTFs are essential for binaural rendering, meaning the reproduction of spatial sound scenes over headphones. Binaural rendering allows virtually placing listeners in an acoustic scene, giving them the impression that they are present and immersed in the virtual acoustic reality. As such, binaural rendering is widely used, for example, in virtual or augmented reality (VR/AR) applications, in the playback of immersive music content, or in acoustic simulations and auralizations~\cite{Roginska2018,Vorlander2020}.

High-quality binaural rendering requires an HRTF set with a dense spatial resolution of $1^\circ - 2^\circ$ according to the minimum audible angle~\cite[Ch.\,2]{Blauert1996} and to avoid artifacts in dynamic binaural reproduction~\cite{Lindau2009a}. Dummy head HRTF sets with high spatial resolution are commonly obtained by sequential measurements~\cite{Gardner1995,Bernschutz2013a,Brinkmann2017c,Li2020a}. However, this is time-consuming, which is why usually speed-optimized methods and equipment are used for measurements of human subjects, such as systems with rotating (semi)circular loudspeaker arcs in combination with signal processing methods that allow continuous rotation of the subject or the arcs~\cite{Algazi2001,Brinkmann2019,Richter2019a,Li2020a}. An appealing approach to reduce the effort, cost, and complexity of measurement systems for individual HRTFs is to measure only a few HRTFs on a sparse spatial sampling grid and then upsample them to a dense HRTF set by interpolation. Conventional measurement systems for acquiring individual HRTFs could thus use fewer loudspeakers and rotate faster. In addition, simplified measurement systems were recently introduced as an alternative to the complex conventional systems~\cite{He2018, Reijniers2020, Bau2021}. Such systems require significantly less equipment and often measure HRTFs for only a few directions on sparse (sometimes even irregular) spatial sampling grids. Here, advanced interpolation methods are essential for upsampling the individual sparse HRTF sets to high-quality dense sets~\cite{Bau2021,Bau2022,Rudzki2022}.

Many interpolation methods for HRTFs have been developed and studied in the last decades (see, for example,~\cite[Ch.\,2.6]{Pike2019} for an overview). The interpolation of HRTFs in the spherical harmonics (SH) domain is popular in spatial audio research and applications~\cite{Evans1998,Porschmann2019,BenHur2019b,Arend2021a}. This global interpolation approach first decomposes the HRTF set into spherical basis functions using the SH transform (also called spherical Fourier transform). The resulting spatially continuous SH representation allows interpolation by applying the inverse SH transform to reconstruct an HRTF for any direction. Another frequently studied global interpolation approach is decomposing an HRTF set based on principal component analysis (PCA) and reconstruction with interpolated PCA weights~\cite{Larcher2000}. Local interpolation algorithms, on the other hand, use a weighted superposition of neighboring HRTFs and differ mainly in the computation of the weights (Barycentric, Natural-Neighbor, Nearest-Neighbor, or Inverse-Distance~\cite{Hartung1999,Gamper2013,Cuevas-Rodriguez2019,BenHur2020,Porschmann2020b}). Choosing the best interpolation algorithm depends on the application. However, perceptually transparent upsampling from a sparse HRTF set is challenging in any case because the HRTF magnitude and phase responses vary greatly with small spatial changes.

The rapid spatial phase changes in HRTFs stem mainly from the varying distance of the ears to the origin of the spherical coordinate system used for HRTF measurements, which is the midpoint between the left and right ear~\cite{Moller1992,Schoerkhuber2018,BenHur2019b}. The distance of the spherical sampling points to the ears changes with the sound incidence direction, adding a broad-band group delay to the HRTFs that changes significantly with the direction and is challenging to interpolate, especially with sparse sampling of the phase response. 

A common approach to this issue is to time-align the HRTFs before interpolation by removing the direction-dependent group delay and reconstructing it after the interpolation by reversing the alignment. For this pre- and postprocessing, several methods have been proposed~\cite{Evans1998,Brinkmann2018a,Porschmann2019,BenHur2019b}, with similar reductions in interpolation errors for all methods~\cite{Arend2021a}. Whereas time alignment generally improves the results of any interpolation method~\cite{Brinkmann2018a,Pike2019,Porschmann2020b,Arend2021a}, the gain in performance varies. For SH interpolation, up to three times fewer directions are required for perceptually transparent interpolation when using alignment~\cite{Arend2021a}.

Current methods for time-aligned interpolation can successfully recover the ITD, which is the main perceptually relevant temporal information in head-related impulse responses (HRIRs, time domain equivalent of the HRTFs), even for extremely sparse sampling grids containing only 6 directions~\cite{Arend2021a}. However, strong position- and frequency-dependent magnitude errors remain that lead to perceived coloration artifacts and loudness/coloration instabilities~\cite{BenHur2019a}, which highlights the need for magnitude-specific pre- and postprocessing. As with the phase response, these errors stem from spatially fast changes in the magnitude response that are challenging to interpolate. Such rapid changes are particularly strong in the contralateral head-shadow region for lateral sound source positions. To the best of our knowledge, these directional changes in HRTF magnitude response have not yet been accounted for by pre- or postprocessing for HRTF interpolation.

To further reduce interpolation errors and thereby the minimum number of HRTFs required for perceptually transparent interpolation, we developed a novel algorithm presented in this paper. The proposed algorithm is based on time-aligned interpolation and introduces an additional postprocessing step for magnitude correction, specifically aiming to reduce remaining coloration artifacts. To derive the magnitude correction, the frequency resolution of the HRTF is reduced to the resolution of the auditory system. The auditory-filtered HRTFs can be interpolated with fewer errors mainly because they are a frequency-smoothed version of the input HRTFs with fewer magnitude changes over space. The interpolation of the auditory-filtered HRTFs is performed in parallel and serves the algorithm as a reference for the magnitude correction, which is applied to the upsampled HRTFs in an additional postprocessing step. Notably, the proposed method does not need additional input to derive the magnitude-correction filters, and it can be combined with any time-alignment and interpolation approach. We refer to the proposed method as Magnitude-Corrected and Time-Aligned interpolation (hereafter abbreviated as MCA).

In the present work, we introduce the MCA method as a general approach for interpolating time-aligned HRTFs. Using SH interpolation, we show in a detailed evaluation based on 96 individual simulated HRTF sets that MCA outperforms conventional interpolation of time-aligned HRTFs, as evidenced by improved magnitude structure and ILDs in the spatially upsampled HRTFs.

\section{Method}

\begin{figure*}[t]
	\centering
	\includegraphics[width=\textwidth]{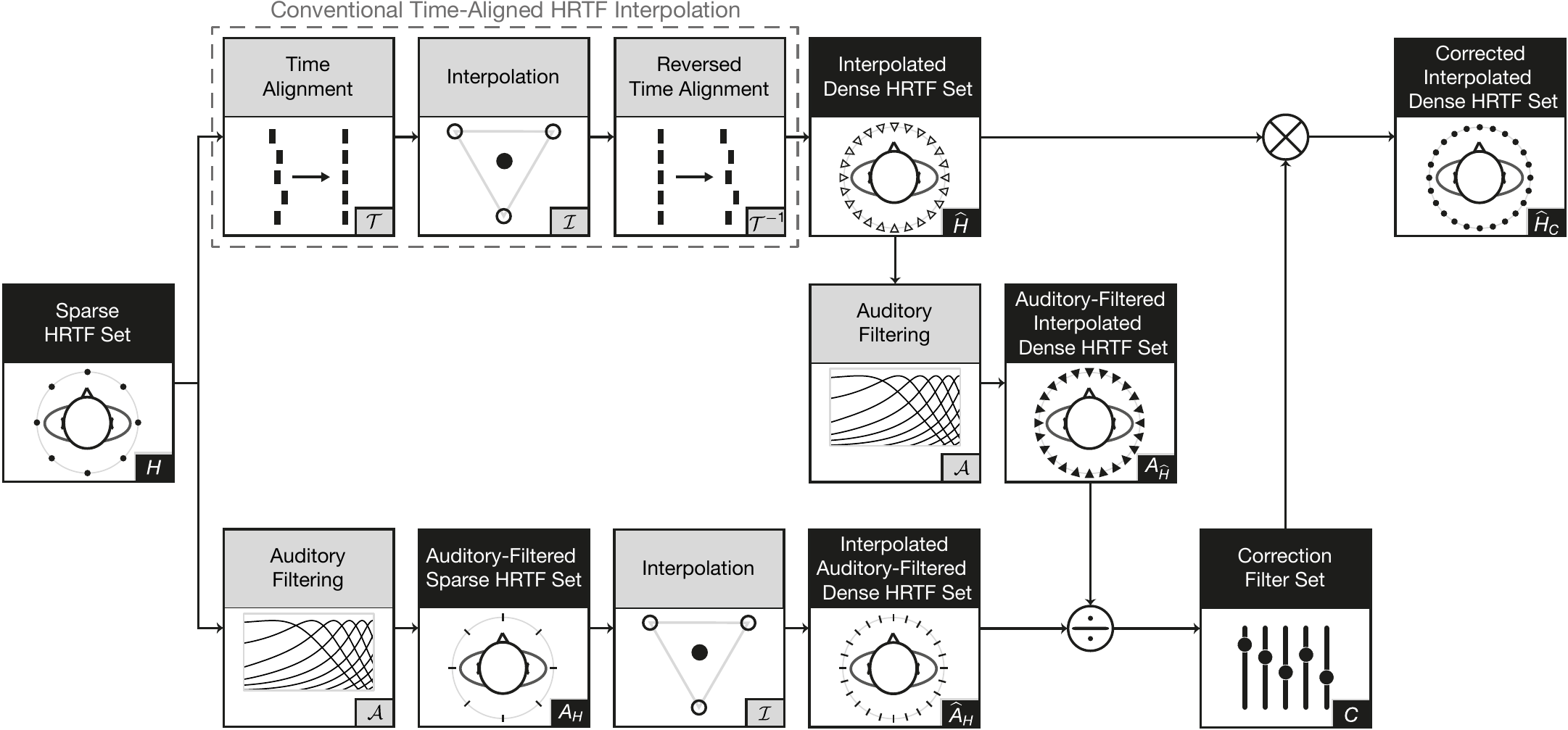}
	\caption{Block diagram of MCA interpolation. A sparse HRTF set is upsampled to a dense set by conventional time-aligned interpolation and then corrected in magnitude using correction filters obtained based on the interpolated auditory-filtered HRTFs, resulting in the final magnitude-corrected dense HRTF set. The blocks with black headers indicate transfer functions, and the blocks with gray headers indicate signal processing operations following the notation used throughout the paper. Conventional time-aligned interpolation is highlighted to illustrate the contributions of MCA interpolation.}
	\label{fig:Blockdiagram}
\end{figure*}

The block diagram in Fig.~\ref{fig:Blockdiagram} shows the basic principle of MCA interpolation and how it is linked to conventional time-aligned interpolation. As the processing is identical for the left and right ear, we did not note the dependency on the ear, the direction, and the frequency in the diagram for clarity. The input $H$, which denotes the sparse HRTF set in the frequency domain, is processed in two ways. On the one hand, a dense HRTF set $\widehat{H}$ is computed by conventional time-aligned HRTF interpolation. This means that the sparse input HRTF set $H$ is first time-aligned, resulting in $H_\mathrm{T} = \mathcal{T}\{H\}$, then interpolated/upsampled to a dense spatial sampling grid, yielding $\widehat{H}_\mathrm{T} = \mathcal{I}\{H_\mathrm{T}\}$, and finally the time alignment is reversed, leading to $\widehat{H} = \mathcal{T}^{-1}\{\widehat{H}_\mathrm{T}\}$, whereby $\widehat{\cdot}$ denotes interpolated data (see, e.g.,~\cite{Arend2021a} for further details on interpolation of time-aligned HRTFs).

On the other hand, $H$ is filtered with auditory Gammatone filters, resulting in the auditory-filtered sparse HRTF set $A_\mathrm{H} = \mathcal{A}\{H\}$, which represents a frequency-smoothed version of the input HRTF set with lower spatial complexity, as magnitude changes over small directional changes are less severe due to averaging the HRTF energy within auditory bands. In the next step, $A_\mathrm{H}$ is interpolated to the same dense target grid, yielding an interpolated auditory-filtered dense HRTF set $\widehat{A}_\mathrm{H} = \mathcal{I}\{A_\mathrm{H}\}$. The set $\widehat{A}_\mathrm{H}$ exhibits smaller magnitude interpolation errors than $\widehat{H}$ and therefore represents the frequency-smoothed target magnitude response of the final upsampled HRTF set. The basic idea of the proposed method is to reduce magnitude interpolation errors in $\widehat{H}$ by applying correction filters derived from the more reliable magnitude structure of $\widehat{A}_\mathrm{H}$.

To obtain the correction filters $C$, auditory Gammatone filters are applied to $\widehat{H}$, yielding the auditory-filtered interpolated dense HRTF set $A_\mathrm{\widehat{H}} = \mathcal{A}\{\widehat{H}\}$. Spectral division of $\widehat{A}_\mathrm{H}$ with $A_\mathrm{\widehat{H}}$ then leads to the frequency-smoothed magnitude response of the correction filters. Through further processing, the correction filters are interpolated to the original frequency resolution of $H$, as Sec.~\,\ref{subsec:Implementation} further details. In a last step, the correction filters are applied to the interpolated HRTFs $\widehat{H}$, resulting in the final interpolated/upsampled and magnitude-corrected dense HRTF set $\widehat{H}_\mathrm{C} = \mathcal{C}\{\widehat{H}\}$.

Note that we intentionally used abstract operators to denote the time alignment $\mathcal{T}\{\cdot\}$, interpolation $\mathcal{I}\{\cdot\}$, and auditory filtering/smoothing $\mathcal{A}\{\cdot\}$ to highlight the generic nature of the proposed algorithm, which can be implemented with different alignment, interpolation, and smoothing approaches.

\subsection{Reference Implementation}\label{subsec:Implementation}

We provide a reference implementation of MCA interpolation as part of the SUpDEq (Spatial Upsampling by Directional Equalization) Matlab toolbox\footnote{Available: \url{https://github.com/AudioGroupCologne/SUpDEq}}, using additional routines from AKtools~\cite{Brinkmann2017c} for filter design and signal processing. MCA interpolation is included in the function \verb|supdeq_interpHRTF|, which provides three different approaches to perform the time alignment $\mathcal{T}\{\cdot\}$ (SUpDEq with different options, Onset-Based Time-Alignment, Phase Correction~\cite{Arend2021a,Porschmann2019,Brinkmann2018a,BenHur2019b}) as well as three approaches for the interpolation $\mathcal{I}\{\cdot\}$ (SH, Natural-Neighbor, Barycentric~\cite{Arend2021a,Porschmann2020b,BenHur2020}). For a detailed description of these procedures, we kindly refer the interested reader to the above references. A demo implementation for spatial upsampling of a sparse HRTF set using MCA interpolation is part of the SUpDEq repository.

One important component of the algorithm is the auditory filtering $\mathcal{A}\{\cdot\}$ of $H$ and $\widehat{H}$. The algorithm uses the filter bank included in the Auditory Toolbox~\cite{Slaney1998} with 41 Gammatone filters $B$ with center frequency $f_c$ distributed on an equivalent rectangular bandwidth scale between 50\,Hz and 20\,kHz. The HRTF in auditory resolution is obtained by filtering the input $X=\{H, \widehat{H}\}$ and accumulating its energy across frequency $f$
\begin{equation}
	\mathcal{A}\{X\} = 10\log_{10}\sum_f B(f_c)|X|^2
\end{equation}
for all $f_c$ and source positions in $X$. 

Designing the correction filters for each direction of the dense target grid is another essential part of the algorithm. The spectral division of $\widehat{A}_\mathrm{H}$ with $A_\mathrm{\widehat{H}}$ yields the initial correction filters in 41 logarithmically spaced frequency bands. These responses are interpolated to linearly spaced frequencies between 0\,Hz and $f_\mathrm{s}/2$ with a resolution of $\Delta f=f_\mathrm{s} / T$, where $f_\mathrm{s}$ is the sampling rate of the HRTFs and $T$ the length of the HRIRs in samples.

Relevant when using SH interpolation, the resulting magnitude responses can optionally be set to 0\,dB below the so-called spatial aliasing frequency $f_\mathrm{A} = N c/(2\pi r_\mathrm{0})$, reasoned because SH interpolation is physically correct below $f_\mathrm{A}$. Here, $c$ denotes the speed of sound, $r_\mathrm{0}$ the head radius, and $N$ corresponds to the SH order of the sparse HRTF set~\cite{Rafaely2015,Bernschutz2016,Arend2021a}. Because $f_\mathrm{A}$ is only an approximation, the correction filters are linearly faded from 0\,dB at $f_\mathrm{A}2^{-1/3}$ to their original value at $f_\mathrm{A}$, thereby introducing a third-octave safety margin below $f_\mathrm{A}$. 

Furthermore, the maximum gain of $C$ can be restricted using soft-limiting applied separately for each frequency bin~\cite[Eq.~(4)]{Giannoulis2012}. The soft-limiting provides a safety measure in case the interpolated auditory-filtered HRTFs $\widehat{A}_\mathrm{H}$ contain errors. Such errors might result in undesired, excessive boosts in $C$, which could lead to audible artifacts or cause problems if the input HRTFs exhibit a low signal-to-noise ratio.

Finally, the compensation filters $C$ are applied by spectral multiplication with $\widehat{H}$, yielding the final magnitude-corrected upsampled dense HRTF set $\widehat{H}_\mathrm{C}$. This step can be realized with a zero- or minimum-phase response for $C$, whereby the latter is derived from the cepstrum of $C$~\cite{Oppenheim2010}.

\section{Technical Evaluation}\label{sec:TechEval}

We evaluated MCA interpolation compared to conventional time-aligned interpolation to investigate the improvements achieved with the proposed magnitude correction. For the evaluation, we used all 96 numerically simulated individual HRTF sets from the HUTUBS database~\cite{Brinkmann2019}, also to demonstrate the general applicability of MCA interpolation. We chose to use simulated instead of measured HRTFs because, as far as we know, there is no freely available dataset of individual HRTFs containing (error-free) data at low elevation angles. When using simulated HRTFs, the measurement-related problem of missing or erroneous data at low elevation angles does not occur, and thus valid full-spherical HRTF sets can be used to evaluate the interpolation method. As a result, the performance of the interpolation algorithm can be examined in isolation, and method-related interpolation errors do not mix with other types of errors due to missing or erroneous data or the approaches used to approximate missing data (see, e.g., \cite{Ahrens2012}). 

In a first application example, we illustrate important processing steps of MCA interpolation and the magnitude structure of the correction filters using one selected HRTF set. In the following, we then discuss results based on all HRTF sets, focusing on spectral differences between the interpolated HRTFs and their reference and to what extent the binaural cues (i.e., the ILDs and ITDs) are affected by the interpolation. Both aspects are also important from a perceptual point of view. The spectral components dominate up/down localization and perceived coloration~\cite{Baumgartner2014}, whereas the binaural cues are essential for left/right localization~\cite{Wightman1992, Blauert1996}.

To generate the sparse input HRTF sets $H$, we spatially resampled the individual full-spherical HRTF sets from the HUTUBS database in the SH domain to Lebedev grids of orders $N = 1 - 15$ ($6 - 350$ sampling points). For time alignment, we applied the SUpDEq method, which achieves the alignment $\mathcal{T}\{H\}$ through a spectral division of $H$ with analytical rigid sphere transfer functions (STFs) for corresponding directions of the sparse grid~\cite{Porschmann2019,Arend2021a}. The reversed alignment $\mathcal{T}^{-1}\{H\}$ after interpolation is performed by a spectral multiplication with STFs for the corresponding directions of the dense target grid. As interpolation method, we used SH interpolation with an SH order corresponding to the respective sparse sampling grid and upsampled to a dense Fliege grid with 900 sampling points ($N=29$). The optimal head radius for each subject required for the subject-specific STFs~\cite{Porschmann2019} was calculated according to Algazi et al.~\cite{Algazi2001a} based on each subject's head width, height, and length. The left and right ear position for the STFs was defined as $\phi = [90^\circ,270^\circ]$ and $\theta = [0^\circ,0^\circ]$ for all subjects. Azimuth angles $\phi = \{0^\circ, 90^\circ, 180^\circ, 270^\circ\}$ denote directions/positions in front, to the left, behind and to the right; elevation angles of $\phi = \{90^\circ, 0^\circ, -90^\circ\}$ directions/positions above, in front, and below. The magnitude-correction filters $C$ were designed as minimum phase filters to avoid the pre-ringing of zero-phase filters. Furthermore, the filters were set to 0\,dB below the respective spatial aliasing frequency $f_\mathrm{A}$ of each individual sparse HRTF set. No soft-limiting was applied to the correction filters throughout this study, as preliminary tests indicated no detrimental effects of unlimited magnitude correction.

\subsection{Application Example}\label{subsec:MethodExample} 
\begin{figure*}[htb]
	\centering
	\hfill
	\subfloat{\includegraphics[width=0.5\textwidth]{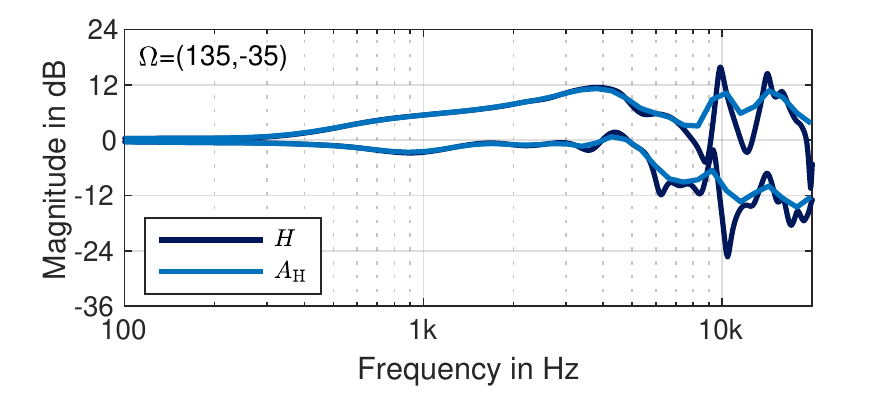}\label{fig:mca_01}}
	\hfill
	\subfloat{\includegraphics[width=0.5\textwidth]{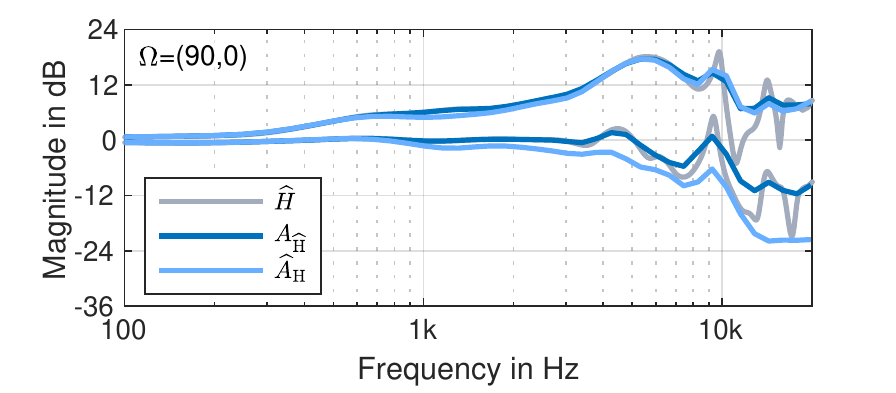}\label{fig:mca_02}}
	\hfill
	\subfloat{\includegraphics[width=0.5\textwidth]{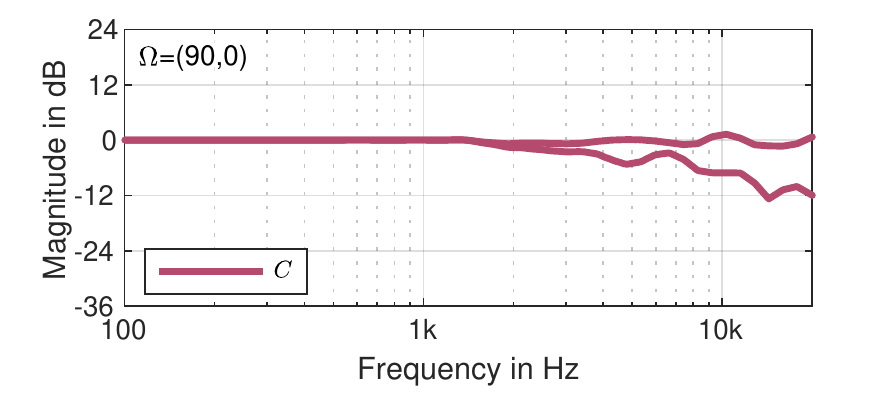}\label{fig:mca_03}}
	\hfill
	\subfloat{\includegraphics[width=0.5\textwidth]{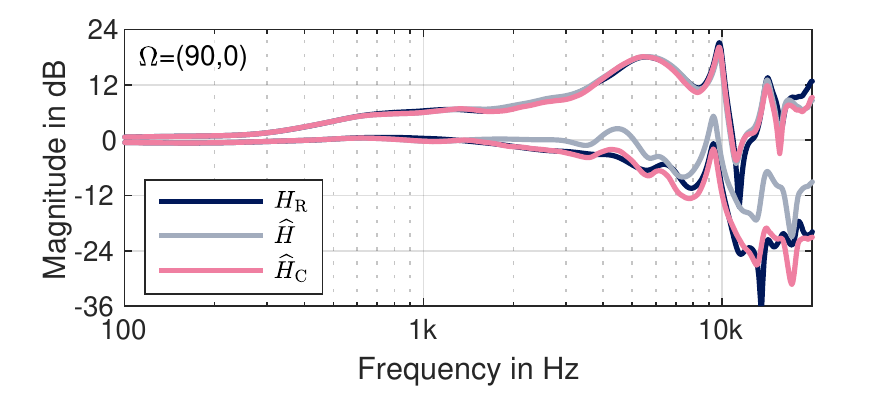}\label{fig:mca_04}}
	\caption{Main processing steps of MCA interpolation for single HRTFs (subject no. 91). For simplicity, the lines for the left and right ear have the same color. The left ear can be distinguished because it has more high frequency energy than the right ear. Top left: HRTF of sparse input set $H$ for $\Omega = (135^\circ,-35^\circ)$ and the respective auditory-filtered HRTF of $A_\mathrm{H}$. Top right: Interpolated HRTF (SUpDEq-processed SH interpolation at $N = 3$) of $\widehat{H}$ for $\Omega = (90^\circ,0^\circ)$ and the respective auditory-filtered interpolated HRTF of $A_\mathrm{\widehat{H}}$ and interpolated auditory-filtered HRTF of $\widehat{A}_\mathrm{H}$. Bottom left: Correction filter $C$ for this specific individual HRTF and direction. Bottom right: HRTF of the reference set $H_\mathrm{R}$, the interpolated HRTF of $\widehat{H}$, and the magnitude-corrected interpolated HRTF of $\widehat{H}_\mathrm{C}$ for this subject and direction.}
	\label{fig:MCAmethod}
\end{figure*}

\begin{figure}[htb]
	\centering
	\includegraphics[width=\columnwidth]{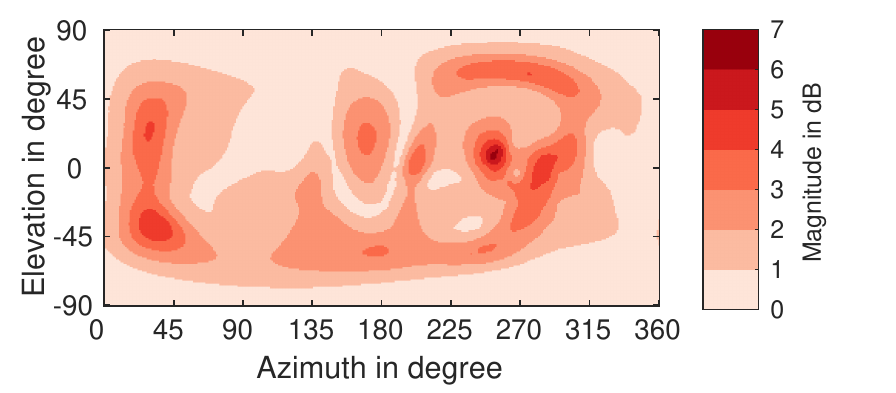}
	\caption{Frequency-averaged absolute magnitude of the left-ear correction filters $C$ over 900 directions of the target Fliege grid. Subject no. 91, SUpDEq-processed SH interpolation at $N = 3$.}
	\label{fig:C_spatial}
\end{figure}

\begin{figure}[htb]
	\centering
	\includegraphics[width=\columnwidth]{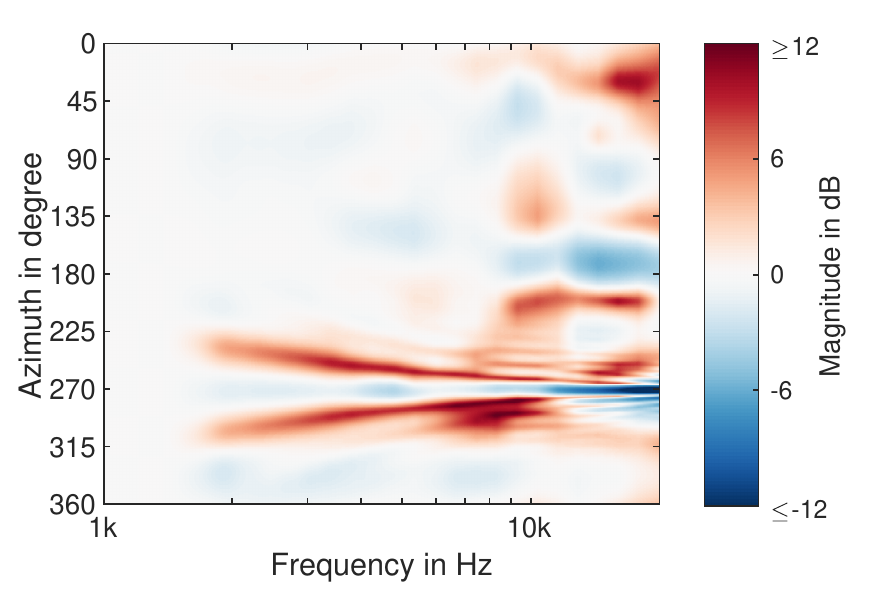}
	\caption{Frequency-dependent signed magnitude of the left-ear correction filters $C$ over the horizontal plane. Subject no. 91, SUpDEq-processed SH interpolation at $N = 3$.}
	\label{fig:C_FREQvsHOR}
\end{figure}

The application example uses the individual HRTF set of subject no. 91 (arbitrarily chosen) from the HUTUBS database, resampled to a sparse HRTF set on a Lebedev grid with 26 sampling points ($N=3$). Calculation of the optimal head radius for subject no. 91 yielded $r_\mathrm{0} = 8.89$\,cm, and the spatial aliasing frequency for this example is $f_\mathrm{A} \approx 1.84$\,kHz.

Figure~\ref{fig:MCAmethod} illustrates the main processing steps for single left and right ear HRTFs. The plot on the top left shows an HRTF of the sparse input set $H$ for the direction $\Omega = (\phi,\theta)$, with $\phi = 135^\circ$ and $\theta = -35^\circ$, as well the respective auditory-filtered HRTF of $A_\mathrm{H}$. The plot on the top right shows an HRTF of $\widehat{H}$ after time-aligned SH interpolation for $\Omega = (90^\circ,0^\circ)$ as well as the auditory-filtered interpolated HRTF $A_{\mathrm{\widehat{H}}}$ and the interpolated auditory-filtered HRTF $\widehat{A}_\mathrm{H}$ for that direction. The plot clearly indicates differences between $A_{\mathrm{\widehat{H}}}$ and $\widehat{A}_\mathrm{H}$, especially at the contralateral right ear, leading to the correction filter $C$ for this specific individual HRTF and direction, as shown in the next plot on the bottom left. The plot on the bottom right shows the reference HRTF $H_\mathrm{R}$ for $\Omega = (90^\circ,0^\circ)$, obtained from the dense full-spherical HRTF set, as well as $\widehat{H}$ for that direction obtained by conventional time-aligned SH interpolation and the magnitude-corrected version $\widehat{H}_\mathrm{C}$ as obtained by MCA interpolation. At the ipsilateral left ear, $\widehat{H}$ and $\widehat{H}_\mathrm{C}$ both closely follow the reference, mainly because time-aligned SH interpolation already provides low interpolation errors for ipsilateral directions~\cite{Porschmann2019,Arend2021a}, and thus almost no magnitude correction is applied. At the contralateral ear, however, differences are more severe, and $\widehat{H}_\mathrm{C}$ matches the reference much better at frequencies above $f_\mathrm{A}$ than $\widehat{H}$. It clearly shows that the magnitude correction in MCA interpolation can efficiently compensate for (broad-band) interpolation errors at the contralateral ear, as evident in $\widehat{H}$ by the magnitude increase at higher frequencies.

To examine which directions are affected the most by the magnitude correction for this particular example, Fig.~\ref{fig:C_spatial} shows the frequency-averaged absolute magnitude of the left-ear correction filters $C$ for all 900 directions of the target Fliege grid. The plot reveals that the magnitude correction operates in large areas along the spherical sampling grid. The correction is particularly strong in the contralateral region, meaning for directions with $\phi=270^\circ \pm 30^\circ$. However, the magnitude correction also applies at the rear for directions with $\phi=180^\circ \pm 25^\circ$ and even more so for directions close to the median plane with $\phi=30^\circ \pm 20^\circ$. In the ipsilateral region at directions with $\phi=90^\circ \pm 30^\circ$, only minor magnitude correction applies because (1) conventional time-aligned SH interpolation already yields good results in this region and (2) similar interpolation errors might occur in $A_{\mathrm{\widehat{H}}}$ and $\widehat{A}_\mathrm{H}$, resulting in only minor corrections. Informal comparisons revealed similar magnitude structures for other subjects from the HUTUBS database, especially in the ipsilateral and contralateral regions. For the directions close to the median plane, however, the correction filters differ among subjects regarding the maximum absolute magnitude, with many of the subjects requiring less correction in this region than in the example shown here.

Fig.~\ref{fig:C_FREQvsHOR} shows the frequency-dependent signed magnitude of left-ear correction filters $C$ in the horizontal plane for this particular example. Interestingly, the figure reveals prominent broadband gains in the contralateral region extending from $240^\circ \leq \phi \leq 260^\circ$ and $280^\circ \leq \phi \leq 300^\circ$. However, in a rather small region at $\phi = 270^\circ \pm 5^\circ$, the filters significantly attenuate, especially at frequencies above 10\,kHz. Furthermore, as intended, the filters converge towards 0\,dB below $f_\mathrm{A}$. The described magnitude structure in the contralateral region with prominent gains and attenuations is similar for all subjects of the HUTUBS database. However, the magnitude behavior of $C$ for other azimuths and frequencies above 10\,kHz is quite individual and varies depending on the subject.

\subsection{Magnitude Errors}\label{sec:MagErrors}
\begin{figure}[h]
	\centering
	\includegraphics[width=\columnwidth]{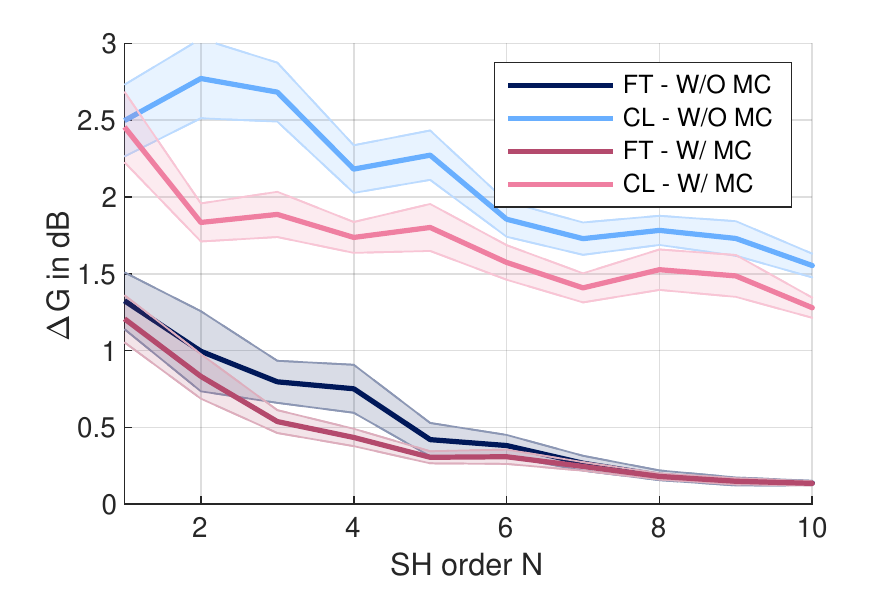}
	\caption{Frequency-, direction-, and subject-averaged left-ear magnitude error $\Delta G$ $\pm$ SD across subjects over SH orders in the frontal (FT) and contralateral (CT) region. SUpDEq-processed SH interpolation without (W/O) and with (W/) magnitude correction (MC).}
	\label{fig:ERBvsN_FrontalContra}
\end{figure}

\begin{figure*}[!ht]
	\centering
	\includegraphics[width=\textwidth]{Figures/ERBvsSPACE_AvgAcrossSubj_SUpDEq_limitMC1.pdf}
	\caption{Frequency- and subject-averaged left-ear magnitude error $\Delta G(\Omega)$ for the SH orders $N = 1 - 5$. SUpDEq-processed SH interpolation without (top) and with (bottom) magnitude correction.}
	\label{fig:ERBvsSPACE}
\end{figure*}

\begin{figure*}[!ht]
	\centering
	\includegraphics[width=\textwidth]{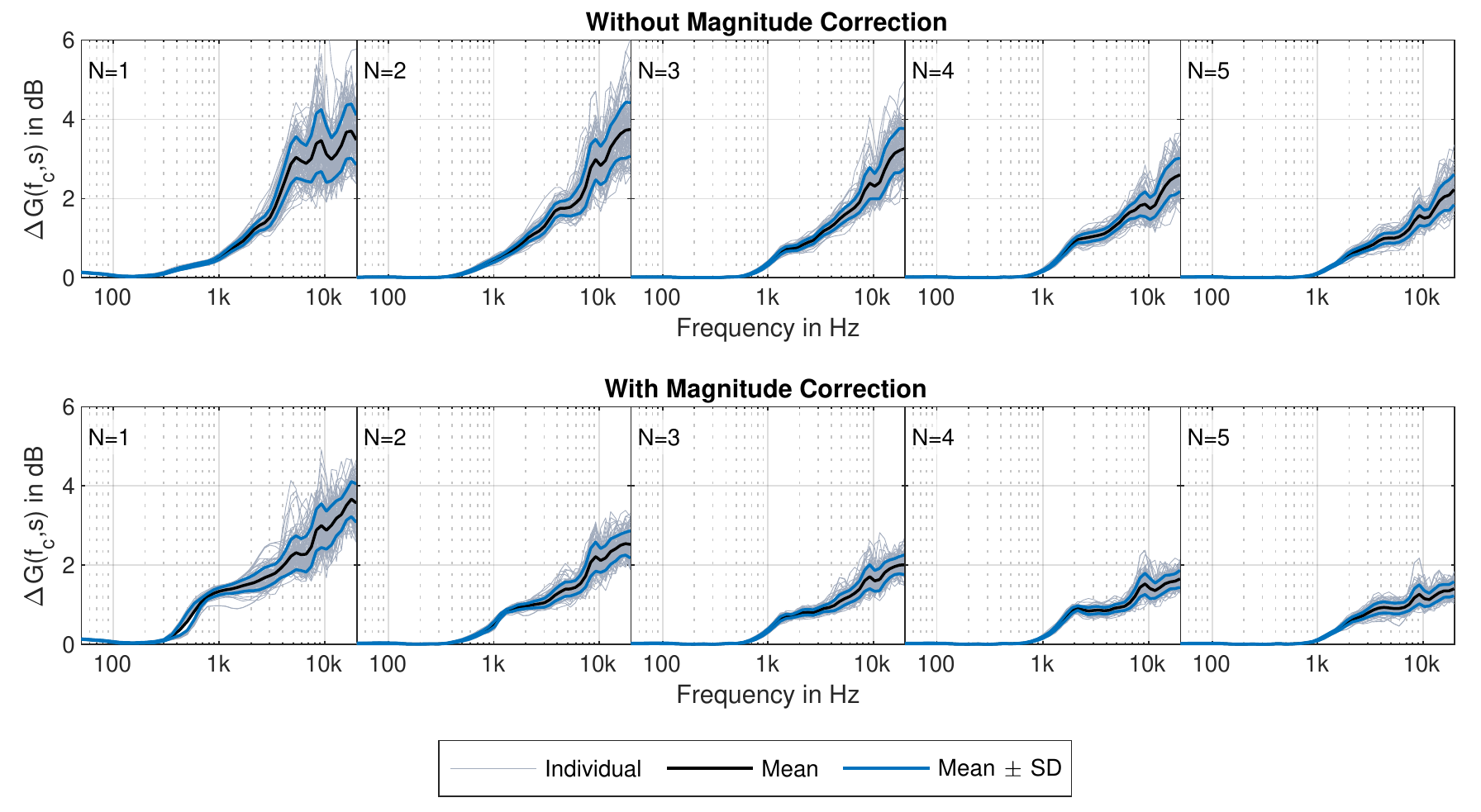}
	\caption{Direction-averaged left-ear magnitude error $\Delta G(f_c,s)$ for the SH orders $N = 1 - 5$. SUpDEq-processed SH interpolation without (top) and with (bottom) magnitude correction.}
	\label{fig:ERBvsFreq_AvgDir}
\end{figure*}

\begin{figure*}[!ht]
	\centering
	\includegraphics[width=\textwidth]{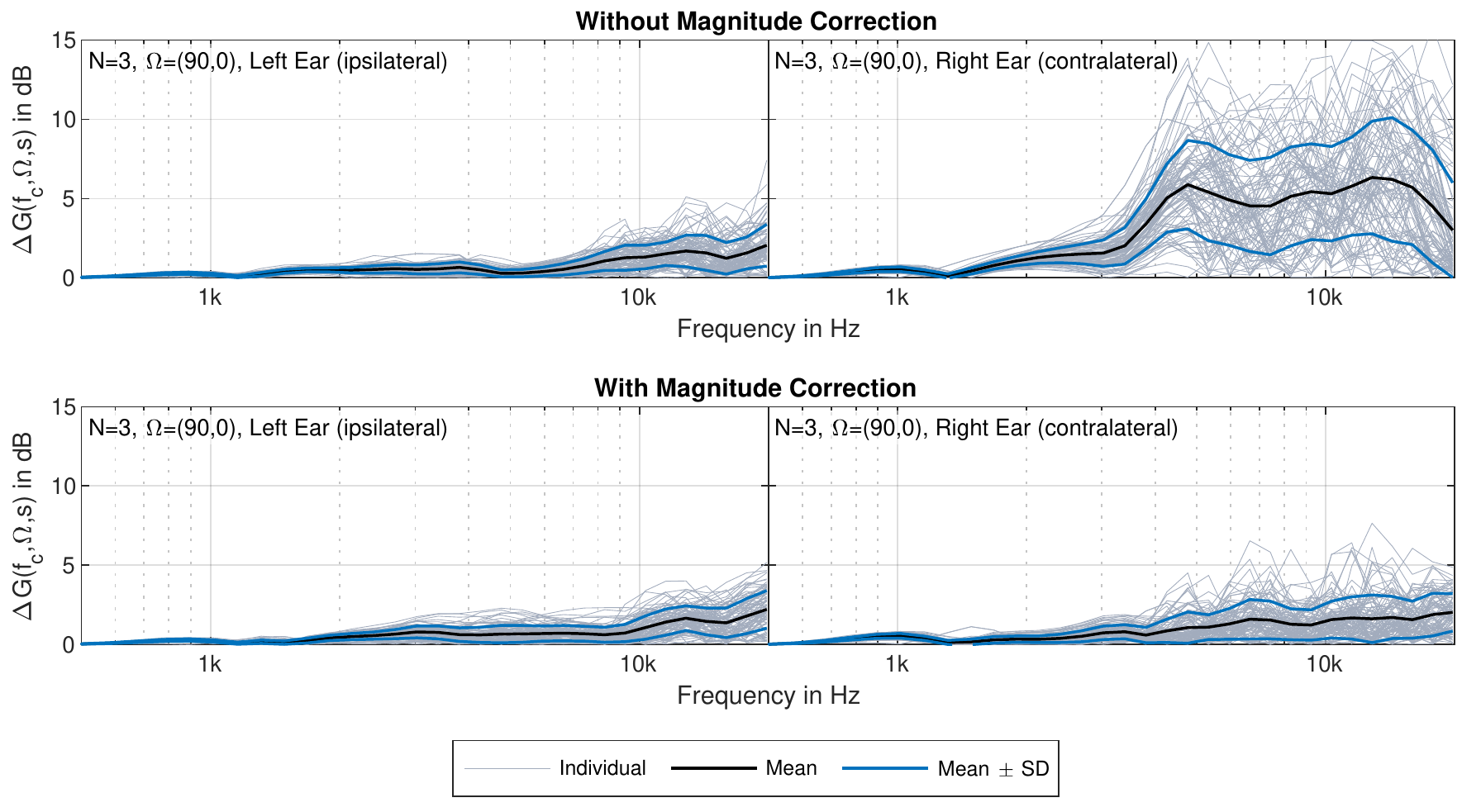}
	\caption{Magnitude error $\Delta G(f_c,\Omega,s)$ at the left and right ear for $\Omega = (90^\circ, 0^\circ)$. SUpDEq-processed SH interpolation at $N = 3$ without (top) and with (bottom) magnitude correction.}
	\label{fig:ERBvsFreq_SingleDirection}
\end{figure*}

This section analyzes the order-dependent magnitude interpolation errors for all 96 individual HRTF sets when applying MCA interpolation or conventional time-aligned interpolation. For each subject $s$ and direction $\Omega$ of the target grid, we determined the magnitude error $\Delta G(f_c,\Omega,s)$ as the absolute energetic difference between the upsampled HRTFs $X=\{\widehat{H},\widehat{H}_\mathrm{C}\}$ and the respective reference HRTFs $H_\mathrm{R}$ in auditory filters with center frequency $f_c$ as
\begin{align}
	\Delta G(f_c,\Omega,s) = \left| 10\log_{10}\frac{\mathcal{A}\{X\}}{\mathcal{A}\{H_\mathrm{R}\}}\right|\ .
	\label{eq:Delta_G}
\end{align}
Averaged errors used in the following are denoted by omitting the corresponding symbol. Thus, $\Delta G(f_c,s)$ describes the subject-specific error averaged across direction, $\Delta G(\Omega)$ represents the error averaged across frequency and subject, and the single value error $\Delta G$ is the average across frequency, direction, and subject.

Figure~\ref{fig:ERBvsN_FrontalContra} shows the left-ear error $\Delta G$ with the standard deviation (SD) across subjects for MCA interpolation (with magnitude correction, $\Delta G$ based on $\widehat{H}_\mathrm{C}$) and conventional time-aligned interpolation (without magnitude correction, $\Delta G$ based on $\widehat{H}$) for frontal and contralateral regions and SH orders up to $N = 10$. The errors were calculated by averaging across directions with $25^\circ$ great circle distance from $\Omega = (0^\circ,0^\circ)$ for the frontal condition and $\Omega = (270^\circ,0^\circ)$ for the contralateral condition. As expected, the errors decrease with increasing SH order and are generally highest at the contralateral ear. However, especially at lower orders $N \leq 5$, the magnitude correction considerably reduces errors in the contralateral region, leading to errors up to 1\,dB smaller than for interpolation without magnitude correction. In the frontal region, the improvements are smaller, with the highest enhancement at $N = \{3,4\}$. Notably, the standard deviation slightly decreases for conditions with magnitude correction, indicating that MCA more consistently reduces interpolation errors across subjects. HRTFs in the ipsilateral region similarly benefit from magnitude correction as HRTFs in the frontal region, which is why we do not show additional error plots for the ipsilateral region. Also, when calculating $\Delta G$ for all directions of the dense Fliege grid, the greatest improvements occur at SH orders $N \leq 5$.

To get a better overview of the spatial distribution of the magnitude errors, Fig.~\ref{fig:ERBvsSPACE} shows the frequency- and subject-averaged left-ear error $\Delta G(\Omega)$ over space for the SH orders $N = 1-5$. Except for $N=1$, where both approaches perform similarly, MCA interpolation reduces the errors for all directions (and each examined HRTF set), generally leading to better interpolation results than conventional time-aligned interpolation. Already at $N = 3$, interpolation errors are below 1\,dB on average in the frontal, ipsilateral, and rear regions. Compared to the HRTFs without magnitude correction, the errors in the contralateral region are spatially less extended and of smaller magnitude. Similar trends show for all SH orders $N \geq 2$.

The frequency-dependent direction-averaged individual left-ear errors $\Delta G(f_c,s)$ in Fig.~\ref{fig:ERBvsFreq_AvgDir} also clearly show the improvements through MCA interpolation. For $N \geq 2$, the errors above the SH-order-dependent spatial aliasing frequency are much smaller in HRTFs with magnitude correction. At frequencies above 10\,kHz, the average error is up to 1.5\,dB lower for MCA interpolation than for conventional time-aligned interpolation. Moreover, with MCA interpolation, the standard deviations are significantly smaller and there are fewer individual outliers, further confirming that magnitude correction improves consistency across individuals.

Last, Fig.~\ref{fig:ERBvsFreq_SingleDirection} provides a closer look at individual frequency-dependent magnitude errors at the ipsi- and contralateral ear for the source position $\Omega = (90^\circ,0^\circ)$. Lateral source positions are of particular interest because the interference patterns they cause at the contralateral ear are especially challenging to interpolate, and the resulting rather strong interpolation errors are clearly audible, as shown in our previous listening experiment~\cite{Arend2021a}. In the specific case of time-aligned SH interpolation of sparse HRTFs, contralateral ear HRTFs usually exhibit strong spatial aliasing artifacts due to the high spatial order of the contralateral magnitude structure caused by strong magnitude changes over small directional changes. In the frequency domain, spatial aliasing appears as the characteristic excessive increase in magnitude at higher frequencies. Figure~\ref{fig:ERBvsFreq_SingleDirection} shows how the magnitude correction of MCA interpolation can significantly reduce these errors at the contralateral ear by attenuating frequencies above $f_\mathrm{A}$ (see also Fig.~\ref{fig:C_FREQvsHOR}). Consequently, the average error at the contralateral ear above 4\,kHz is about 4\,dB lower with MCA interpolation than with time-aligned interpolation only, probably rendering artifacts inaudible in many cases. In line with previous observations, the standard deviation decreases when applying magnitude correction. The errors for the ipsilateral ear are similar regardless of the interpolation method applied. Here, conventional time-aligned interpolation already performs well, and the magnitude correction cannot reduce the errors much further.

\subsection{Binaural Cues}\label{sec:BinCues}
This section examines SH-order-dependent ILD and ITD errors for all 96 individual interpolated HRTF sets. For each subject $s$, we determined the broadband ILD as the energetic ratio of the left and right ear HRIRs. We then calculated the absolute ILD errors for 360 directions $\Omega$ in the horizontal plane (i.e., $0^\circ \leq \phi \leq 359^\circ, \theta = 0^\circ$) as
\begin{align}
	\Delta \mathrm{ILD}(\Omega,s) = \biggm\lvert
	\underbrace{10\log_{10} \frac{\sum x_\mathrm{l}^2}{\sum x_\mathrm{r}^2}}_{\text{ILD of $x$}}  -
	\underbrace{10\log_{10} \frac{\sum h_\mathrm{R,l}^2}{\sum h_\mathrm{R,r}^2}}_{\text{ILD of $h_\mathrm{R}$}}
	\biggm\lvert\ ,
	\label{eq:Delta_ILD}
\end{align} 
where $x = \{\widehat{h}, \widehat{h}_\mathrm{C}\}$ denotes the interpolated HRIRs and $h_\mathrm{R}$ the respective reference HRIRs, and the subscripts $l$ and $r$ indicate the left and right ear, respectively. 

To determine the ITD errors, we estimated the low-frequency ITD as the differences in time-of-arrival (TOA) between the left and right ear. We then calculated the absolute ITD errors in the horizontal plane for each subject $s$ as 
\begin{align}
	\Delta \mathrm{ITD}(\Omega,s) = |
	\underbrace{[\mathcal{O}(x_\mathrm{l}) - \mathcal{O}(x_\mathrm{r})]}_{\text{ITD of $x$}}
	-
	\underbrace{[(\mathcal{O}(h_\mathrm{R,l}) - \mathcal{O}(h_\mathrm{R,r})]}_{\text{ITD of $h_\mathrm{R}$}}
	|\ ,
	\label{eq:Delta_ITD}
\end{align}
where $\mathcal{O}$ denotes the abstract operator for TOA estimation in HRIRs. In the present case, the TOAs were estimated using onset detection with a threshold of $-10$\,dB in relation to the maximum values of the 10 times upsampled and low-pass-filtered HRIRs (8th order Butterworth filter, cut-off frequency 3\,kHz, see~\cite{Andreopoulou2017a}).

\begin{figure*}[!ht]
	\centering
	\includegraphics[width=\textwidth]{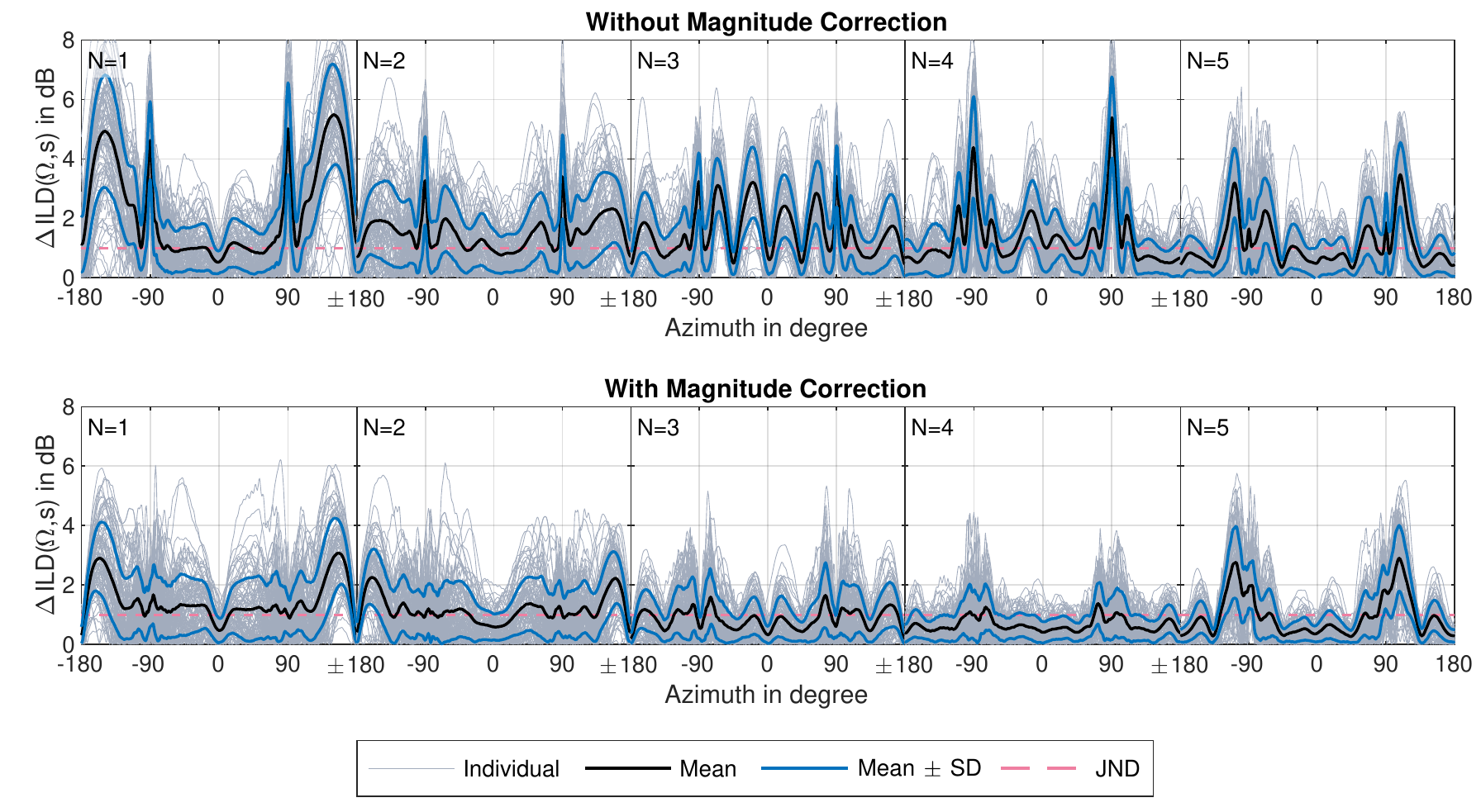}
	\caption{Horizontal plane ILD errors $\Delta \mathrm{ILD}(\Omega,s)$ for the SH orders $N = 1-5$. SUpDEq-processed SH interpolation without (top) and with (bottom) magnitude correction.}
	\label{fig:ILDvsPosition}
\end{figure*}

\begin{figure*}[!ht]
	\centering
	\includegraphics[width=\textwidth]{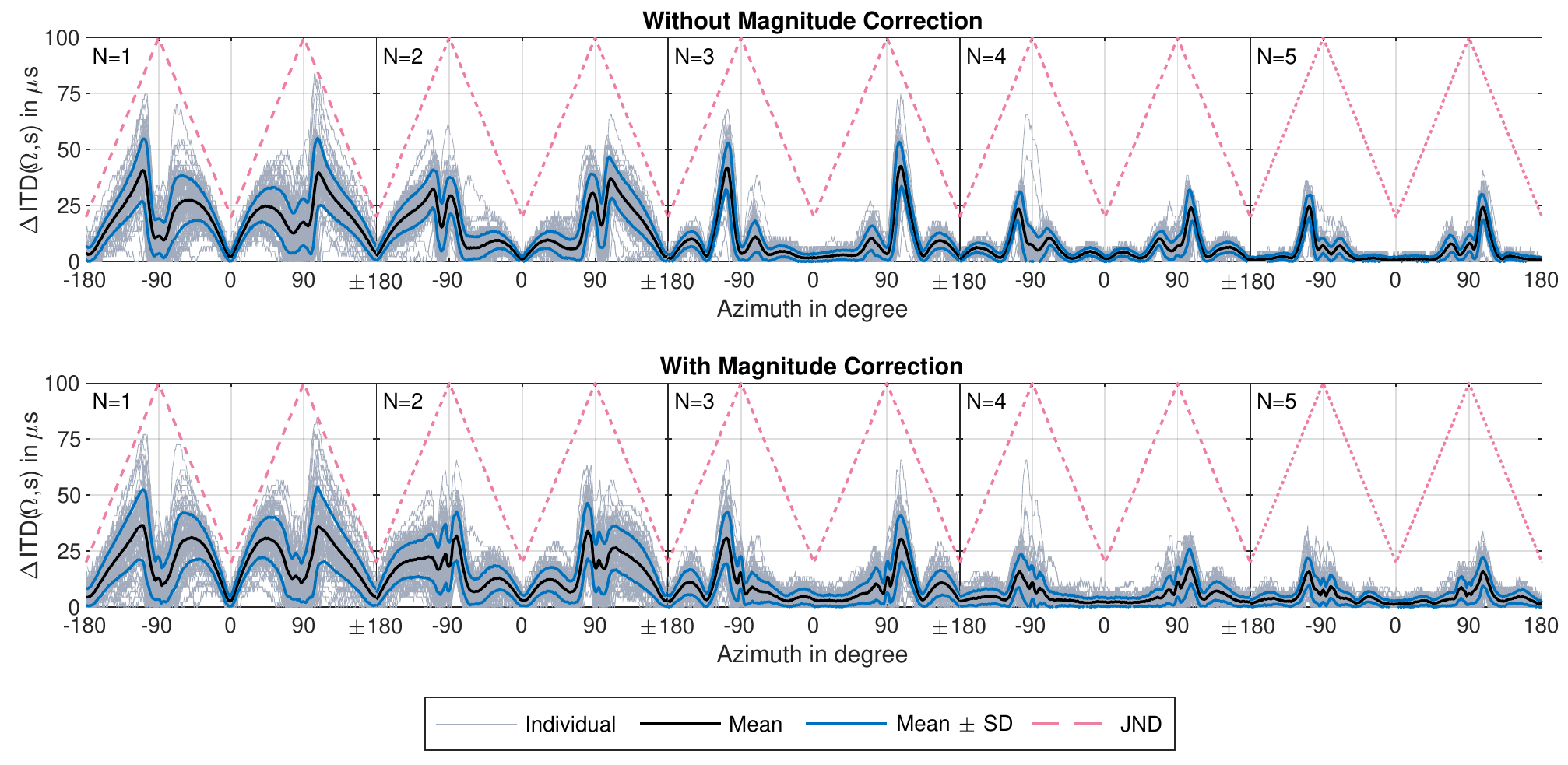}
	\caption{Horizontal plane ITD errors $\Delta \mathrm{ITD}(\Omega,s)$ for the SH orders $N = 1-5$. SUpDEq-processed SH interpolation without (top) and with (bottom) magnitude correction.}
	\label{fig:ITDvsPosition}
\end{figure*}

Figure~\ref{fig:ILDvsPosition} shows the individual ILD errors $\Delta \mathrm{ILD}(\Omega,s)$ in the horizontal plane along with the mean and standard deviation across subjects for SH orders $N = 1-5$. To indicate the errors' perceptual importance, the dashed lines show the direction-independent broadband just-noticeable difference (JND) of 1\,dB~\cite{Blauert1996}. MCA interpolation results in considerably lower ILD errors than time-aligned interpolation. Most notably, the magnitude correction drastically reduces ILD errors for lateral and rear source positions (i.e., $|\phi| \gtrsim 90^\circ$), demonstrating the logical consequence of the independent magnitude correction for the left and right ear HRTFs: reduced (broadband) ILD errors. Even at $N = 1$, where the overall magnitude error with MCA interpolation was not clearly lower (cf. Sec.~\ref{sec:MagErrors}), the ILD errors are significantly decreased, with a reduction in the mean maximum error of more than 2\,dB. With magnitude correction, already at $N = 3$, the mean error is in the range of the JND for all horizontal directions, and at $N = 4$, also the standard deviation across subjects approaches the JND, and the mean maximum error is decreased by more than 3\,dB, indicating perceptually correct ILDs for most directions and approximately 68\,\% of the subjects. Surprisingly, at $N = 5$, the ILD errors at lateral positions for the magnitude-corrected HRTFs increase slightly again by about $1-2$\,dB in average, although one would expect the errors to decrease with increasing SH order. We assume that these are sampling-scheme-dependent artifacts and that by using other input grids instead of the employed Lebedev grid, the errors would distribute differently in the horizontal plane or decrease with increasing SH order as expected.

Last, Fig.~\ref{fig:ITDvsPosition} shows the individual horizontal plane ITD errors $\Delta \mathrm{ITD}(\Omega,s)$ along with the mean and standard deviation across subjects for SH orders $N = 1-5$. The direction-dependent JND was calculated by linearly interpolating between broadband JNDs of 20\,$\mu$s at $\phi = \{0^\circ,\pm180^\circ\}$ and 100\,$\mu$s at $\phi = \pm90^\circ$~\cite{Mossop1998}. Overall, the ITD errors are below the JND for all SH orders, subjects, and directions examined, regardless of whether magnitude correction was applied, indicating no perceptual impairments or localization errors in the interpolated HRTFs related to ITDs. The low errors are due to the applied SUpDEq time-alignment method, which is the most accurate analytical alignment procedure because it considers wave-based effects that affect the low-frequency ITDs~\cite{Benichoux2016,Arend2021a}. Importantly, the error plots reveal that the additional filtering of the interpolated HRTFs for magnitude correction has no negative impact on their broadband ITDs. Rather, the ITDs are slightly improved at $N \geq 3$ when using MCA interpolation. 

\section{Discussion}
For the spatial upsampling of sparse HRTF sets, various interpolation approaches and pre- and postprocessing methods have been developed to reduce interpolation errors. Most current methods perform similarly well, but magnitude interpolation errors, especially in contralateral regions, remain challenging and still require relatively dense sampling for perceptually transparent interpolation \cite{Porschmann2020b,Arend2021a}. To further reduce these interpolation errors and consequently decrease the minimum number of HRTFs required for perceptually transparent interpolation, we developed MCA interpolation, which combines a magnitude correction with the conventional time-aligned interpolation approaches.

In this paper, we introduced the MCA method as a generic approach for interpolating HRTFs, applying any of the recent time-alignment or interpolation approaches together with the proposed magnitude correction. In the technical evaluation, we examined the MCA method based on SH interpolation in conjunction with SUpDEq processing for time alignment, employing all 96 individual simulated HRTF sets from the HUTUBS database. In an application example, we illustrated the main processing steps of MCA interpolation and described in more detail the frequency- and direction-dependent properties of the magnitude correction filters $C$ applied to the interpolated HRTFs $\widehat{H}$ in order to obtain the final magnitude-corrected dense HRTF set $\widehat{H}_\mathrm{C}$. The analysis revealed prominent boosts and attenuations by the correction filters, especially in the contralateral region. As found in further informal analyses, the magnitude structure of $C$ in the contralateral region illustrated in the application example for one subject is typical for all subjects in the HUTUBS database. 

The technical evaluation further showed that, compared to sole time-aligned SH interpolation with SUpDEq processing, the MCA method (or, in particular, the included magnitude correction) significantly reduces magnitude errors in the interpolated HRTFs for all tested subjects and thus improves the quality of the upsampled HRTF sets. The improvements are most pronounced at lower SH orders $N \leq 5$ and contralateral/rear regions, as conventional methods usually already perform well in frontal and ipsilateral regions and at higher SH orders. The analysis further revealed that magnitude-corrected and time-aligned SH interpolation at $N = 3$ has similar error levels as conventional time-aligned SH interpolation at $N = 6$. These findings indicate that the MCA method requires only about one-third of input HRTFs compared to conventional time-aligned SH interpolation in order to achieve similar interpolation results, that is, 16 (N = 3) instead of 49 directions (N = 6).

The results for the analysis of binaural cues (ILDs and ITDs) in interpolated HRTFs revealed additional benefits of MCA interpolation. As a consequence of the independent magnitude correction for the left and right ear, MCA interpolation leads to significantly lower ILD errors, especially for lateral and rear source positions, which, on average across subjects, are already in the JND range at $N = 3$. In contrast, sole time-aligned SH interpolation requires higher SH orders $N \geq 6$ for similarly low error levels. The ITD error analysis confirmed results from previous studies that SH interpolation with SUpDEq time alignment exhibits negligible ITD errors below the JND even at $N = 1$~\cite{Arend2021a}. The additional filtering of the HRTFs for magnitude correction has no negative influence on the ITDs, but even slightly reduces ITD errors. In summary, the technical evaluation clearly shows that the magnitude correction improves the quality of upsampled HRTFs in several respects, but never degrades the interpolation results compared to interpolation with time alignment only.

The described technical improvements also significantly enhance the perceptual quality of the upsampled HRTFs. In informal listening allowing source and head movements, we found much better direction-dependent loudness- and timbre stability~\cite{BenHur2019a} in HRTFs upsampled with the MCA method than with solely time-aligned SH interpolation, which is a result of the improved magnitude structure and ILDs. For moving noise sound sources, the magnitude correction appears to render differences between upsampled HRTFs and their reference already inaudible at much lower SH orders ($N \approx 3-5$) than without magnitude correction. As for the critical lateral source positions~\cite{Arend2021a}, the magnitude correction significantly reduces the errors at the contralateral ear (cf. Fig.~\ref{fig:ERBvsFreq_SingleDirection}), also drastically improving the perceptual quality of (static) lateral sound sources. To confirm these informal results, we plan to conduct listening experiments to determine perceptual thresholds for indistinguishability between interpolated and reference HRTFs as a function of the SH order, just as in one of our recent studies~\cite{Arend2021a}, as well as qualitative listening tests to compare different methods for time-aligned interpolation with and without magnitude correction.

The proposed algorithm can be modified or parameterized at specific points, and we compared different configurations during development. In particular, we investigated whether using fractional-octave smoothing instead of frequency smoothing with auditory filters improves the interpolation results, but we found no relevant differences between the two smoothing methods. Further, we examined whether soft-limiting should generally be used and to what extent it affects the interpolation results. Here, we found that soft-limiting the filters to 6\,dB with a smooth knee (e.g., $3-6$\,dB) yields similar results as when no limiting is applied at all, suggesting that, in general, few correction filters have really strong level boosts and soft-limiting is not strictly necessary in most cases.

With the proposed method, remaining magnitude errors in auditory bands according to Eq.~\eqref{eq:Delta_G} can be almost entirely attributed to interpolation errors in the interpolated auditory-filtered dense HRTF set $\widehat{A}_\mathrm{H}$, which serves as a reference for the magnitude correction, and the subsequent construction of the magnitude correction filters. The remaining errors thus indicate that even the frequency-smoothed sparse input HRTFs still have such high spatial complexity that common HRTF interpolation methods cannot produce error-free results. As a consequence, errors in the final upsampled HRTF set $\widehat{H}_\mathrm{C}$ would automatically be smaller if the interpolation results of the frequency-smoothed sparse input HRTF had fewer errors. In future research, we thus aim to find either (1) an improved interpolation method for frequency-smoothed HRTFs that produces fewer interpolation errors and/or (2) a more compact and perceptually valid representation of the input HRTF to further reduce the spatial complexity. That said, the broadband magnitude correction of the MCA method is by design incapable of perfectly reconstructing detailed notch structures occurring especially in contralateral HRTFs (cf. Fig.~\ref{fig:MCAmethod}). Spectral notches, however, are usually less salient~\cite{Moore1989}, and the signal level at the contralateral ear is much lower and of less importance to sound source localization~\cite{Baumgartner2014}. We thus assume that inaccurate reconstruction of the notch structure in the contralateral region of upsampled HRTFs has negligible perceptual impact. 

In addition to the ideas for future work outlined above, we plan to evaluate the MCA method in follow-up studies with various current interpolation (e.g., SH, Barycentric, Natural-Neighbor) and time-alignment approaches (e.g., SUpDEq, Onset-Based Time-Alignment, Phase Correction). In preliminary studies, we found that the magnitude correction compensates for the differences between the various approaches and always improves interpolation results to a similar extent as presented in this paper, but these results need to be confirmed in a more detailed analysis. Furthermore, we plan to evaluate MCA interpolation for irregular~\cite{Bau2022} and/or incomplete~\cite{Ahrens2012} sparse sampling grids. Also for such cases, we assume that the magnitude correction in MCA interpolation can significantly reduce interpolation errors and improve the quality of the upsampled HRTFs.

\section{Conclusion}
In this paper, we presented magnitude-corrected and time-aligned (MCA) interpolation, a novel approach for spatial upsampling of HRTFs. To the best of our knowledge, it is the first approach that explicitly targets a reduction of magnitude interpolation errors by combining time-aligned interpolation with post-interpolation magnitude correction based on an analysis and processing of the HRTFs in auditory bands. The evaluation of the algorithm showed the expected reduction in magnitude interpolation errors and the potential to further reduce the minimum number of HRTFs required for perceptually transparent interpolation. 

\section*{Acknowledgments}
This work was funded by the German Research Foundation (DFG WE 4057/21-1).

\bibliographystyle{IEEEtran}
%\bibliography{Audio&Acoustics.bib}
\normalsize
\bibliography{Audio&Acoustics.bib}
%

%%%%%%%%%%%%%%   Bibliography   %%%%%%%%%%%%%%
%\normalsize
%\bibliography{references}

%%%%%%%%%%%%  Supplementary Figures  %%%%%%%%%%%%
%\clearpage

%%%%%%%%%%%%%%%%   End   %%%%%%%%%%%%%%%%
%\end{multicols}  % Method B for two-column formatting (doesn't play well with line numbers), comment out if using method A
\end{document}